\documentclass[showpacs, showkeys]{revtex4}

\usepackage[small]{caption}
\usepackage{amsmath,amsfonts,amscd,amssymb,epsf, epsfig,mathrsfs}\usepackage{graphicx}

\newcommand{\pa}{\partial}
\newcommand{\vep}{\varepsilon}
\begin{document}
\title{Fermionic Casimir interaction in cylinder-plate and cylinder-cylinder geometries}

\author{L. P. Teo}
 \email{LeePeng.Teo@nottingham.edu.my}
 \affiliation{Department of Applied Mathematics, Faculty of Engineering, University of Nottingham Malaysia Campus, Jalan Broga, Semenyih, 43500, Selangor Darul Ehsan, Malaysia.}
\begin{abstract}In this work, we consider the Casimir effect due to massless fermionic fields in the presence of long cylinders. More precisely, we consider the interaction between a cylinder parallel to a plate, between two parallel cylinders outside each other, and between a cylinder lying parallelly inside another cylinder. We derive the explicit formulas for the Casimir interaction energies and compute the leading and the next-to-leading order terms of the small separation asymptotic expansions. As expected, the leading order terms coincide with the proximity force approximations. We compare the results of the next-to-leading  order terms of different quantum fields, and show that our results support the ansatz of derivative expansions.

\end{abstract}
 \maketitle

\section{Introduction}

Casimir effect is one of the most interesting predictions of quantum field theory that has been verified experimentally. In \cite{57}, Casimir proposed that there exists a force between two parallel perfectly conducting plates due to the vacuum fluctuations of the electromagnetic field confined between the plates. The idea of Casimir energy is natural. In quantum theory, the ground state energy of a simple harmonic oscillator is not zero, but is equal to $ \hbar \omega/2$, where $\omega$ is the angular frequency of the associated oscillator. A quantum field can be considered as the superposition of infinitely many simple harmonic oscillators with different frequencies. This led Casimir to define the Casimir energy to be
\begin{align}\label{eq3_27_1}
E_{\text{Cas}}=\sum  \frac{\hbar\omega}{2},
\end{align}the sum of the ground state energies of the quantum field. This is an infinite sum that needs regularization. However, without the presence of boundaries or external conditions, the Casimir effect would not be manifested. The Casimir effect is most interesting in the presence of two objects, such as parallel plates. In principle, after subtracting the self energies, there should be some finite amount of energy left, which would create interaction between the two objects.

The first experiment that successfully verified the presence of Casimir effect appeared near the end of the 20th century \cite{58}.
This has stimulated another surge in the research activities in Casimir effect, especially in conjunction with the development of nanotechnology. Under the same reasoning, the definition of Casimir energy \eqref{eq3_27_1} works not only for electromagnetic fields as originally proposed by Casimir, but for any quantum fields. However, for fermionic fields, one should add a minus sign in front of the formula due to the different spin and statistical behavior of fermionic fields. In fact, such definitions of Casimir energies have been  used to compute the Casimir effect of two parallel plates, and the Casimir self energies of spheres, cylinders, etc, in the second half of the 20th century. From the point of view of statistical physics, such a definition is natural since the Casimir energy so defined appears as the zero temperature part of the free energy.

Even though there is intensive research in Casimir effect since 1980s, for a long time, it is not clear how to compute the Casimir interaction between two objects, except by using approximations. Around 2006, various groups of researchers simultaneously tackled this problem for some particular geometries using quantum field theory methods such as Green's functions, path integrals, wave expansions, etc, that can be more or less categorized as multiple scattering approach \cite{21, 22, 23, 24, 25, 27, 28, 14, 15, 30, 31, 32, 33,70} or mode summation approach \cite{34, 35, 36}. The general method for arbitrary objects has been synthesized in \cite{26} for scalar fields and in \cite{29} for electromagnetic fields. Both of these papers approach the problem using multiple scattering formalism. In \cite{1}, we used mode summation approach to interpret the formulas derived in \cite{26, 29}. An advantage of our formalism is that it is not restricted to scalar fields or electromagnetic fields, and it is also not restricted to (3+1)-dimensional Minkowski spacetime. For example, we have used the formalism in \cite{1} to compute the sphere-plate and sphere-sphere Casimir interaction in $(D+1)$-dimensional Minkowski spacetime in \cite{17} and \cite{18}.

The studies of Casimir effect of fermionic fields can be dated back to 1980s. Before the end of the 20th century, there are a few works that considered Casimir effect of fermionic fields \cite{37,6,38,5}, but this is a relatively small number compared to the research works in Casimir effect of scalar fields and electromagnetic fields. One of the possible reasons is that  ferminoic fields are relatively harder to deal with. Nevetheless, after entering the 21st century, Casimir effect of fermionic fields has started to attract   attention and there are a number of works in this area \cite{39, 40, 41, 42, 43, 44, 45, 46, 47, 48, 55, 49, 71,72, 50, 51, 52, 53, 56, 54}, with some applications to carbon nanotubes.

In \cite{59}, we considered the fermionic Casimir interaction between two spheres and derived the  small separation asymptotic behaviors. For application to carbon nanotubes, it is  natural to consider the cylindrical geometries. In this work, we consider the fermionic Casimir effect between a cylinder and a plate, and between two parallel cylinders. For parallel cylinders, we consider both possible cases: one is where the two cylinders are outside each other, and one is where one cylinder is inside the other. We derive the explicit formulas for the Casimir interaction energies and compute the small separation asymptotic behaviors. The results are compared to the results of other quantum fields. We also use our results to stipulate the ansatz of derivative expansions proposed in \cite{60}.

\section{The Casimir interaction energy}
\subsection{Plane waves and cylindrical waves}
In this work, we consider the  Casimir interaction between a cylinder and a plate, and between two cylinders due to the vacuum fluctuations of a massless Dirac field $\psi$
 which satisfies the equation
\begin{equation}
i\gamma^{\mu}\nabla_{\mu}\psi=0.
\end{equation}
Here
$
\nabla_{\mu}=\pa_{\mu}+\Gamma_{\mu},
$ and $\Gamma_{\mu}$ is the spin connection.

 On the boundaries of the cylinders or plate, we impose the MIT bag boundary conditions:
\begin{equation}\label{eq5_16_2}
(1+i\gamma^{\mu}n_{\mu})\psi\Bigr|_{\text{boundary}}=0.
\end{equation}
To derive the Casimir interaction energy, we use the formalism we developed in \cite{1}.

First we need to solve the equation of motion \eqref{eq5_16_2} in rectangular and cylindrical coordinates.

For  cylinders, we   align them so that their axes of symmetry are parallel to the $z$-direction. Then using the cylindrical coordinates
\begin{equation}
x=\rho\cos\varphi,\quad y=\rho\sin\varphi,\quad z=z,
\end{equation}  a cylinder of radius $R$ can be described as $\rho=R$ if its axis of symmetry is the $z$-axis.

When considering the cylinder-plate interaction, we will assume that the plate is given by $x=L$, where $L>R$, so that the cylinder is parallel to the plate.
The plane waves are then parametrized by $(k_y, k_z)$. The fermionic waves can be divided into positive energy modes and negative energy modes, as well as regular waves and outgoing waves, each has two families: They can be written as
\begin{equation}
\psi^{(\pm), *}_{k_yk_z, j}=A^{(\pm), *}_{k_yk_z, j}e^{-i\text{sgn}_*k_xx+ik_yy+ik_zz\mp i\omega t},
\end{equation}where
\begin{equation}\begin{split}
A_{k_yk_z, 1}^{(\pm), *}= &\begin{pmatrix} 1 \\0  \\ \displaystyle \pm \frac{k_z}{k}\\ \displaystyle\pm \frac{-\text{sgn}_* k_x+ik_y}{k}\end{pmatrix},\\
A_{k_yk_z, 2}^{(\pm), *}=& \begin{pmatrix} 0 \\1  \\ \mp \displaystyle\frac{\text{sgn}_*k_x+ik_y}{k}\\  \displaystyle\mp\frac{k_z}{k}\end{pmatrix}.
\end{split}\end{equation}
Here $\displaystyle k=\frac{\omega}{c}$, $*$ = reg or out, $\text{sgn}_{\text{reg}}=1$, $\text{sgn}_{\text{out}}=-1$ and $$k_x=\sqrt{k^2-k_y^2-k_z^2}.$$

In cylindrical coordinates, the fermionic waves are parametrized by $m$ and $k_z$, where $\displaystyle m=\pm\frac{1}{2}, \pm\frac{3}{2}, \pm\frac{5}{2}, \ldots$,
\begin{align}
\psi_{mk_z, j}^{(\pm),*}=\mathcal{C}_m^{*}B_{mk_z, j}^{(\pm),*}e^{ik_zz\mp i\omega t},
\end{align}with
\begin{equation}\begin{split}
B_{mk_z, 1}^{(\pm),*}=& \begin{pmatrix}
   f_{m-\frac{1}{2}}^*(k_{\perp}\rho)e^{i\left(m-\frac{1}{2}\right)\varphi}\\ 0\\ \pm  \displaystyle  \frac{k_z}{k}  f_{m-\frac{1}{2}}^*(k_{\perp}\rho)e^{i\left(m-\frac{1}{2}\right)\varphi }\\\pm     \displaystyle \frac{ik_{\perp}}{k}   f_{m+\frac{1}{2}}^*(k_{\perp}\rho)e^{i\left(m+\frac{1}{2}\right)\varphi }
\end{pmatrix},\\
B_{mk_z, 2}^{(\pm),*}=& \begin{pmatrix}0\\
 -i  f_{m+\frac{1}{2}}^*(k_{\perp}\rho)e^{i\left(m+\frac{1}{2}\right)\varphi } \\ \mp    \displaystyle \frac{k_{\perp}}{k}  f_{m-\frac{1}{2}}^*(k_{\perp}\rho)e^{i\left(m-\frac{1}{2}\right)\varphi }\\\pm \displaystyle \frac{i    k_z}{k}    f_{m+\frac{1}{2}}^*(k_{\perp}\rho)e^{i\left(m+\frac{1}{2}\right)\varphi }
\end{pmatrix}.
\end{split}\end{equation}
Here
$$k_{\perp}=\sqrt{k^2-k_z^2},$$
\begin{align*}
f_n^{\text{reg}}(z)=J_n(z),\quad f_n^{\text{out}}=H_n^{(1)}(z),
\end{align*}and
\begin{align}
\mathcal{C}_m^{\text{reg}}= i^{-m+\frac{1}{2}},\quad \mathcal{C}_m^{\text{out}}=\frac{\pi}{2}i^{m+\frac{1}{2}}
\end{align}are normalization constants introduced to facilitate the change to imaginary frequencies.

\subsection{The Casimir interaction energy between a cylinder and a plate}
Assume that the cylinder has radius $R$, length $H$ and its axis of symmetry is the $z$-axis. Let the plate be described by $x=L$, with center at $O'=(L,0,0)$ and dimensions $H\times H$.
Let $\mathbf{x}'=\mathbf{x}-\mathbf{L}$, $\mathbf{L}=L\mathbf{e}_x$.

In the region between the cylinder and the plate, the Dirac fields can be represented in two ways: one is in terms of the cylindrical coordinate system centered at $O$:
\begin{equation}\begin{split}
\psi^{(\pm)}(\mathbf{x},t) =&H\int_{-\infty}^{\infty} d\omega \sum_{m=-\infty}^{\infty}\left(\sum_{j=1,2}a_j^{(\pm), mk_z} \psi_{mk_z,j}^{(\pm),\text{reg}}(\mathbf{x},\omega)
 +\sum_{j=1, 2}b_j^{(\pm), mk_z} \psi_{mk_z,j}^{(\pm),\text{out}}(\mathbf{x},\omega) \right);\end{split}
\end{equation}and one is in terms of the rectangular coordinate system centered at $O'$:
\begin{equation}\begin{split}
\psi^{(\pm)}(\mathbf{x}',t) =&H^2\int_{-\infty}^{\infty} d\omega \int_{-\infty}^{\infty}\frac{dk_y}{2\pi}\int_{-\infty}^{\infty}\frac{dk_z}{2\pi} \left(
\sum_{j=1, 2}c_j^{(\pm), k_yk_z }\psi^{(\pm), \text{reg}}_{k_yk_z, j }(\mathbf{x}',\omega) +\sum_{j=1,2}d_j^{(\pm), k_yk_z }\psi^{(\pm), \text{out}}_{k_yk_z, j }(\mathbf{x}',\omega) \right).\end{split}
\end{equation}These two representations are related by translation matrices $\mathbb{V}$ and $\mathbb{W}$:
\begin{equation}\label{eq3_27_3}\begin{split}
\begin{pmatrix}\psi^{(\pm), \text{reg}}_{k_yk_z, 1 }(\mathbf{x}',\omega)\\
\psi^{(\pm), \text{reg}}_{k_yk_z, 2 }(\mathbf{x}',\omega)\end{pmatrix}=&\sum_{n=-\infty}^{\infty}\begin{pmatrix} V^{(\pm), 11}_{m, k_y}& V^{(\pm), 21}_{m, k_y}\\
V^{(\pm), 12}_{m, k_y} & V^{(\pm), 22}_{m, k_y}\end{pmatrix}\begin{pmatrix}
 \psi_{mk_z,1}^{(\pm),\text{reg}}(\mathbf{x},\omega)\\\psi_{mk_z,2}^{(\pm),\text{reg}}(\mathbf{x},\omega)\end{pmatrix}, \\
\begin{pmatrix}\psi^{(\pm), \text{out}}_{mk_z, 1 }(\mathbf{x},\omega)\\\psi^{(\pm), \text{out}}_{mk_z, 2 }(\mathbf{x},\omega)\end{pmatrix}
=&H  \int_{-\infty}^{\infty}\frac{dk_y}{2\pi}  \begin{pmatrix}
W^{(\pm), 11}_{ k_y, m}& W^{(\pm), 21}_{ k_y, m}\\ W^{(\pm), 12}_{ k_y, m} & W^{(\pm), 22}_{ k_y, m}\end{pmatrix}\begin{pmatrix}\psi_{k_yk_z,1}^{(\pm),\text{out}}(\mathbf{x}',\omega)\\\psi_{k_yk_z,2}^{(\pm),\text{out}}(\mathbf{x}',\omega)\end{pmatrix},
\end{split}\end{equation}so that
\begin{equation}\begin{split}
\begin{pmatrix}
a_1^{(\pm),mk_z}\\a_2^{(\pm),mk_z}
\end{pmatrix}=&H\int_{-\infty}^{\infty}\frac{dk_y}{2\pi} \begin{pmatrix}V_{m, k_y}^{(\pm), 11}  & V_{ m,k_y}^{(\pm), 12}  \\
 V_{m,k_y}^{(\pm), 21}  & V_{m,k_y}^{(\pm), 22}  \end{pmatrix}\begin{pmatrix}
c_1^{(\pm),k_yk_z}\\c_2^{(\pm),k_yk_z}
\end{pmatrix},\\
\begin{pmatrix}
d_1^{(\pm),k_yk_z}\\d_2^{(\pm),k_yk_z}
\end{pmatrix}=&\sum_{m=-\infty}^{\infty}\begin{pmatrix}W_{k_y, m}^{(\pm),11}  & W_{k_y, m}^{(\pm),12}  \\
W_{k_y, m}^{(\pm),21}  & W_{k_y, m}^{(\pm),22} \end{pmatrix}\begin{pmatrix}
b_1^{(\pm),mk_z}\\b_2^{(\pm),mk_z}
\end{pmatrix}.\end{split}
\end{equation}
The boundary conditions on the cylinder give a relation of the form
\begin{align}
\begin{pmatrix}
b_1^{(\pm), mk_z}\\b_2^{(\pm), mk_z}
\end{pmatrix}=-\mathbb{T}_{mk_z}^{(\pm)}\begin{pmatrix}
a_1^{(\pm), mk_z}\\a_2^{(\pm), mk_z}
\end{pmatrix}.
\end{align}In imaginary frequency, $\omega=i\xi$, $k=i\kappa$, $\gamma=\sqrt{\kappa^2+k_z^2}$,
\begin{equation}\label{eq3_26_2}\begin{split}
\mathbb{T}_{m k_z}^{(\pm)}=&\pm i\frac{1}{\kappa R\left(K_{m-\frac{1}{2}}^2(\gamma R)+K_{m+\frac{1}{2}}^2(\gamma R)\right)}
 \begin{pmatrix}\displaystyle  -1\mp i\mathscr{A} &\displaystyle
\frac{ik_z}{\gamma}\\ \displaystyle \frac{ik_z}{\gamma}& \displaystyle 1 \mp i\mathscr{A}\end{pmatrix},
\end{split}\end{equation}
where
\begin{align}\label{eq3_26_4}
\mathscr{A}=\kappa R \left(K_{m-\frac{1}{2}}(\gamma R)I_{m-\frac{1}{2}}(\gamma R)- K_{m+\frac{1}{2}}(\gamma R)I_{m+\frac{1}{2}}(\gamma R)\right).
\end{align}
The boundary conditions on the plate gives a relation of the form
\begin{align}
\begin{pmatrix}
c_1^{(\pm),k_yk_z}\\c_2^{(\pm),k_yk_z}
\end{pmatrix}=-\widetilde{\mathbb{T}}^{(\pm)}_{k_yk_z}\begin{pmatrix}
d_1^{(\pm),k_yk_z}\\d_2^{(\pm),k_yk_z}
\end{pmatrix},
\end{align}where
\begin{align}
\widetilde{\mathbb{T}}^{(\pm)}_{k_yk_z}= \mp\frac{ i}{\kappa}\begin{pmatrix} \displaystyle  \sqrt{\gamma^2+k_y^2 }+k_y  &\displaystyle  ik_z \\
\displaystyle - ik_z & \displaystyle \displaystyle  \sqrt{\gamma^2+k_y^2 }-k_y \end{pmatrix}.
\end{align}
 For the translation matrices $\mathbb{V}$ and $\mathbb{W}$ defined by \eqref{eq3_27_3},
using techniques introduced in \cite{1}, we find that
\begin{align}
\mathbb{V}_{m,k_y}=\begin{pmatrix}1&0\\0 & \displaystyle \frac{\sqrt{\gamma^2+k_y^2 }+k_y}{\gamma} \end{pmatrix}\left(\frac{\sqrt{\gamma^2+k_y^2 }+k_y}{\gamma}\right)^{m-\frac{1}{2}}  e^{-\sqrt{\gamma^2+k_y^2 }L},
\end{align}
 \begin{align}
\mathbb{W}_{k_y,m}=\frac{ \pi}{H}\begin{pmatrix}1&0\\0 & \displaystyle -\frac{\sqrt{\gamma^2+k_y^2 }+k_y}{\gamma} \end{pmatrix}\left(\frac{\sqrt{\gamma^2+k_y^2 }+k_y}{\gamma}\right)^{m-\frac{1}{2}} \frac{e^{-\sqrt{\gamma^2+k_y^2 }L}}{\sqrt{\gamma^2+k_y^2 }}.
\end{align}
As discussed in \cite{1}, the Casimir interaction energy is then given by
\begin{equation}\label{eq3_26_1}
E_{\text{Cas}}
=-\frac{\hbar H}{2\pi}\int_0^{\infty} d\xi \int_{-\infty}^{\infty}\frac{ dk_z}{2\pi}\sum_{+, -}\text{Tr}\,\ln  \left(\mathbb{I}-\mathbb{M}^{(\pm)}(i\xi)\right),
\end{equation}where
\begin{equation}\begin{split}
\mathbb{M}_{m, m'}^{(\pm)}=&\mathbb{T}_{mk_z}^{(\pm)}H\int_{-\infty}^{\infty}\frac{dk_y}{2\pi}\mathbb{V}_{m,k_y}^{(\pm)}\widetilde{\mathbb{T}}_{k_yk_z}^{(\pm)}\mathbb{W}_{k_y,m'}^{(\pm)}
\\  =&\mp i\frac{\pi}{\kappa}\mathbb{T}_{mk_z}^{(\pm)} \begin{pmatrix}\gamma& \displaystyle -i k_z \\\displaystyle -ik_z & \displaystyle -\gamma\end{pmatrix} \int_{-\infty}^{\infty}\frac{dk_y}{2\pi}\left(\frac{\sqrt{\gamma^2+k_y^2 }+k_y}{\gamma}\right)^{m+m' } \frac{e^{-2\sqrt{\gamma^2+k_y^2 }L}}{\sqrt{\gamma^2+k_y^2 }}. \end{split}
\end{equation}
The integral can be computed explicitly to give
\begin{equation}\begin{split}
\mathbb{M}_{m, m'}^{(\pm)}=&\mp \frac{ i}{ \kappa}\mathbb{T}_{mk_z} \begin{pmatrix}\gamma& \displaystyle -i k_z \\\displaystyle -ik_z & \displaystyle -\gamma\end{pmatrix}
K_{m+m'}\left(2\gamma L\right).
\end{split}\end{equation}
\subsection{The Casimir interaction energy between two cylinders  }
Consider two cylinders of length $H$ and  radii $R_A$ and $R_B$ respectively. The axis of symmetry of the cylinders are given respectively by $x=y=0$ and $x=L, y=0$, both parallel to the $z$ axis.

We consider two scenarios:
\begin{enumerate}
\item[I.] The two cylinders are outside each other. In this case, $L>R_A+R_B$ and $d=L-R_A-R_B$ is the distance between the two cylinders.

\item[II.] The cylinder of radius $R_A$ is inside the cylinder of radius $R_B$. In this case, $L<R_B-R_A$ and $d=R_B-R_A-L$ is the distance between the two cylinders.
\end{enumerate}

In the region between the two cylinders, the Dirac fields can be represented in two ways: one is in terms of the cylindrical coordinate system centered at $O$:
\begin{equation}\begin{split}
\psi^{(\pm)}(\mathbf{x},t) =&H\int_{-\infty}^{\infty} d\omega \sum_{m=-\infty}^{\infty}\left(\sum_{j=1,2}a_j^{(\pm), mk_z} \psi_{mk_z,j}^{(\pm),\text{reg}}(\mathbf{x},\omega)
 +\sum_{j=1, 2}b_j^{(\pm), mk_z} \psi_{mk_z,j}^{(\pm),\text{out}}(\mathbf{x},\omega) \right);\end{split}
\end{equation}and one is in terms of the cylindrical  coordinate system centered at $O'$:
\begin{equation}\begin{split}
\psi^{(\pm)}(\mathbf{x},t) =&H\int_{-\infty}^{\infty} d\omega  \sum_{m'=-\infty}^{\infty}\left(
\sum_{j=1, 2}c_j^{(\pm), m'k_z }\psi^{(\pm), \text{reg}}_{m'k_z, j }(\mathbf{x}',\omega) +\sum_{j=1,2}d_j^{(\pm), m'k_z }\psi^{(\pm), \text{out}}_{m'k_z, j }(\mathbf{x}',\omega) \right).\end{split}
\end{equation}
The two representations are related by translation matrices. In case that the two cylinders are outside each other,
\begin{equation}\begin{split}
\begin{pmatrix}\psi^{(\pm), \text{out}}_{m'k_z, 1 }(\mathbf{x}',\omega)\\
\psi^{(\pm), \text{out}}_{m'k_z, 2 }(\mathbf{x}',\omega)\end{pmatrix}=&\sum_{m=-\infty}^{\infty}\begin{pmatrix} U^{(\pm), 11}_{m, m'}& U^{(\pm), 21}_{m, m'}\\
U^{(\pm), 12}_{m, m'} & U^{(\pm), 22}_{m, m'}\end{pmatrix}\begin{pmatrix}
 \psi_{mk_z,1}^{(\pm),\text{reg}}(\mathbf{x},\omega)\\\psi_{mk_z,2}^{(\pm),\text{reg}}(\mathbf{x},\omega)\end{pmatrix}, \\
\begin{pmatrix}\psi^{(\pm), \text{out}}_{mk_z, 1 }(\mathbf{x},\omega)\\
\psi^{(\pm), \text{out}}_{mk_z, 2 }(\mathbf{x},\omega)\end{pmatrix}=&\sum_{m'=-\infty}^{\infty}\begin{pmatrix} \widetilde{U}^{(\pm), 11}_{m', m}& \widetilde{U}^{(\pm), 21}_{m', m}\\
\widetilde{U}^{(\pm), 12}_{m', m} & \widetilde{U}^{(\pm), 22}_{m', m}\end{pmatrix}\begin{pmatrix}
 \psi_{m'k_z,1}^{(\pm),\text{reg}}(\mathbf{x}',\omega)\\\psi_{m'k_z,2}^{(\pm),\text{reg}}(\mathbf{x}',\omega)\end{pmatrix}.
\end{split}\end{equation}
In case one cylinder is inside the other,
\begin{equation}\begin{split}
\begin{pmatrix}\psi^{(\pm), \text{reg}}_{m'k_z, 1 }(\mathbf{x}',\omega)\\
\psi^{(\pm), \text{reg}}_{m'k_z, 2 }(\mathbf{x}',\omega)\end{pmatrix}=&\sum_{m=-\infty}^{\infty}\begin{pmatrix} V^{(\pm), 11}_{m, m'}& V^{(\pm), 21}_{m, m'}\\
V^{(\pm), 12}_{m, m'} & V^{(\pm), 22}_{m, m'}\end{pmatrix}\begin{pmatrix}
 \psi_{mk_z,1}^{(\pm),\text{reg}}(\mathbf{x},\omega)\\\psi_{mk_z,2}^{(\pm),\text{reg}}(\mathbf{x},\omega)\end{pmatrix}, \\
\begin{pmatrix}\psi^{(\pm), \text{out}}_{mk_z, 1 }(\mathbf{x},\omega)\\
\psi^{(\pm), \text{out}}_{mk_z, 2 }(\mathbf{x},\omega)\end{pmatrix}=&\sum_{m'=-\infty}^{\infty}\begin{pmatrix}W^{(\pm), 11}_{m', m}& W^{(\pm), 21}_{m', m}\\
W^{(\pm), 12}_{m', m} &W^{(\pm), 22}_{m', m}\end{pmatrix}\begin{pmatrix}
 \psi_{m'k_z,1}^{(\pm),\text{out}}(\mathbf{x}',\omega)\\\psi_{m'k_z,2}^{(\pm),\text{out}}(\mathbf{x}',\omega)\end{pmatrix}.
\end{split}\end{equation}

Using the   method in \cite{1}, one finds that
\begin{equation}\begin{split}
\mathbb{U}_{m,m'}^{(\pm)}=&(-1)^{m'-\frac{1}{2}}K_{m'-m}(\gamma L)\mathbb{I},
\\
\mathbb{U}_{m',m}^{(\pm)}=&(-1)^{m'-\frac{1}{2}}K_{m'-m}(\gamma L)\mathbb{I},\\
\mathbb{V}_{m,m'}^{(\pm)}=&(-1)^{m'-m}I_{m-m'}(\gamma L)\mathbb{I},\\
\mathbb{W}_{m',m}^{(\pm)}=&(-1)^{m-m'}I_{m-m'}(\gamma L)\mathbb{I}.
\end{split}\end{equation}As in the case of electromagnetic fields, we find that the translation matrices are all equal to a scalar times the identity matrix.

When the two cylinders are outside each other, solving the boundary conditions on the two cylinders give the $\mathbb{T}^{(\pm)}_{mk_z}$ matrix as in \eqref{eq3_26_2}.
The Casimir interaction energy is given by
\begin{equation}\label{eq3_26_3}
E_{\text{Cas}}
=-\frac{\hbar H}{2\pi}\int_0^{\infty} d\xi \int_{-\infty}^{\infty}\frac{ dk_z}{2\pi}\sum_{+, -}\text{Tr}\,\ln  \left(\mathbb{I}-\mathbb{M}^{(\pm)}(i\xi)\right),
\end{equation}
with
\begin{equation}\begin{split}
\mathbb{M}_{m,m'}^{(\pm)}=&\mathbb{T}_{mk_z}^{(\pm)}(R_A)\sum_{m''}\mathbb{U}_{m,m''}^{(\pm)}\mathbb{T}_{m''k_z}^{(\pm)}(R_B)\widetilde{\mathbb{U}}_{m'',m'}^{(\pm)}\\
=&\mathbb{T}_{mk_z}^{(\pm)}(R_A)\sum_{m''}K_{m''-m}(\gamma L)\mathbb{T}_{m''k_z}^{(\pm)}(R_B)K_{m''-m'}(\gamma L).
\end{split}\end{equation}

When the cylinder with radius $R_A$ is inside the cylinder with radius $R_B$, the boundary condition on the cylinder with radius $R_A$ still give the same $\mathbb{T}_{nk_z}^{(\pm)}$ as given by \eqref{eq3_26_2}. However, the boundary conditions on the cylinder with radius $R_B$ gives
 \begin{equation}\begin{split}
\begin{pmatrix}
c_1^{(\pm), m'k_z}\\c_2^{(\pm), m'k_z}
\end{pmatrix}=-\widetilde{\mathbb{T}}_{m'k_z}^{(\pm)}\begin{pmatrix}
d_1^{(\pm), m'k_z}\\d_2^{(\pm), m'k_z}
\end{pmatrix},
\end{split}\end{equation}
where
\begin{equation}\begin{split}
\widetilde{\mathbb{T}}^{mk_z}=& \pm i\frac{1}{\kappa R_B\left(I_{m-\frac{1}{2}}^2(\gamma R_B)+I_{m+\frac{1}{2}}^2(\gamma R_B)\right)}
 \begin{pmatrix}\displaystyle  -1\mp i\mathscr{A} &\displaystyle
\frac{ik_z}{\gamma}\\ \displaystyle \frac{ik_z}{\gamma}& \displaystyle 1 \mp i\mathscr{A}\end{pmatrix},
\end{split}\end{equation}with $\mathscr{A}$ given by \eqref{eq3_26_4}.
The Casimir interaction energy is then  given by the same expression \eqref{eq3_26_3} but with
\begin{equation}\begin{split}
\mathbb{M}_{m,m'}^{(\pm)}=&\mathbb{T}_{mk_z}^{(\pm)}(R_A)\sum_{m''}\mathbb{V}_{m,m''}^{(\pm)}\widetilde{\mathbb{T}}_{m''k_z}^{(\pm)}(R_B) \mathbb{W}_{m'',m'}^{(\pm)}\\
=&\mathbb{T}_{mk_z}^{(\pm)}(R_A)\sum_{m''}I_{m''-m}(\gamma L)\widetilde{\mathbb{T}}_{m''k_z}^{(\pm)}(R_B)I_{m''-m'}(\gamma L).
\end{split}\end{equation}

\section{Small separation asymptotic behavior }
Casimir effect will be most significant when the separation between the objects is small. However, the computation of the small separation asymptotics of the Casimir interaction energy is often a tedious problem. Nonetheless, a systematic method has been developed in a series of papers \cite{14,15,7,8,16,9,11,12,13,10,17,18}. Making the substitution
$$k_z=u\cos\alpha,\quad \kappa=u\sin\alpha$$in \eqref{eq3_26_1} and \eqref{eq3_26_3}, we find that
\begin{equation}\begin{split}
E_{\text{Cas}}
=&-\frac{\hbar cH }{ 4\pi^2} \int_{0}^{\infty} udu\int_0^{\pi} d\alpha \text{Tr}\,\sum_{+, -}\ln  \left(\mathbb{I}-\mathbb{M}^{(\pm)} \right).
\end{split}\end{equation}
Expanding the logarithm and trace, we have
\begin{equation}\begin{split}
E_{\text{Cas}}
=&\frac{\hbar cH }{ 4\pi^2}\sum_{+, -} \sum_{s=0}^{\infty}\frac{1}{s+1}\int_{0}^{\infty} udu\int_0^{\pi} d\alpha \sum_{m_0 }
\sum_{m_1 }\ldots\sum_{m_s }\text{tr}\,\left(\mathbb{M}^{(\pm)}_{m_0m_1}\mathbb{M}^{(\pm)}_{m_1m_2}\ldots \mathbb{M}^{(\pm)}_{m_sm_0}\right).
\end{split}\end{equation}Here the trace tr is the trace over $2\times 2 $ matrices.

In the following, we are going to discuss the different cases separately.
\subsection{The cylinder-plate case}
In the cylinder-plate case,
let
$$\omega=Ru,\quad\vep=\frac{d}{R}=\frac{L}{R}-1.$$Then
\begin{equation}\begin{split}
\mathbb{M}_{m_i, m_{i+1}}^{(\pm)}=&  \frac{1}{\omega\sin^2\alpha \left(K_{m_i-\frac{1}{2}}^2( \omega)+K_{m_i+\frac{1}{2}}^2(  \omega)\right)}\begin{pmatrix}-1\mp i \mathscr{B}_i\sin\alpha & i\cos\alpha \\ i\cos\alpha & 1\mp i\mathscr{B}_i\sin\alpha \end{pmatrix} \begin{pmatrix}1& \displaystyle -i  \cos\alpha\\\displaystyle -i \cos\alpha & \displaystyle -1\end{pmatrix}
K_{m_i+m_{i+1}}\left(2\omega(1+\vep)\right)\\
=&   \frac{1}{\omega\sin^2\alpha \left(K_{m_i-\frac{1}{2}}^2( \omega)+K_{m_i+\frac{1}{2}}^2(  \omega)\right)}\begin{pmatrix}-\sin^2\alpha\mp i \mathscr{B}_i\sin\alpha & \mp \mathscr{B}_i \sin\alpha\cos\alpha \\ \mp \mathscr{B}_i \sin\alpha\cos\alpha& -\sin^2\alpha\pm i\mathscr{B}_i\sin\alpha \end{pmatrix}
K_{m_i+m_{i+1}}\left(2\omega(1+\vep)\right),
\end{split}\end{equation}where
$$\mathscr{B}_i=\omega \left(K_{m_i-\frac{1}{2}}(\omega)I_{m_i-\frac{1}{2} }(\omega)- K_{m_i+\frac{1}{2}}(\omega)I_{m_i+\frac{1}{2}}(\omega)\right).$$
Let
\begin{gather*}
m_0=m, \quad m_i=m+n_i,\\
\omega=\frac{m\sqrt{1-\tau^2}}{\tau}.
\end{gather*}
Then the main contribution to the Casimir interaction energy comes from $m\sim \vep^{-1}$, $n_i\sim \vep^{-\frac{1}{2}}$.

Using the Debye asymptotic expansions of modified Bessel functions:
 \begin{equation}\label{eq3_26_6}\begin{split}
I_{\nu}(\nu z)\sim & \frac{1}{\sqrt{2\pi \nu}}\frac{e^{\nu\eta(z)}}{(1+z^2)^{\frac{1}{4}}}\left(1+\frac{u_1(t(z))}{\nu}+\ldots\right),\\
K_{\nu}(\nu z)\sim &\sqrt{\frac{\pi}{ 2 \nu}}\frac{e^{-\nu\eta(z)}}{(1+z^2)^{\frac{1}{4}}}\left(1-\frac{u_1(t(z))}{\nu}+\ldots\right),
\end{split}\end{equation}where
\begin{equation}\begin{split}
\eta(z)=&\sqrt{1+z^2}+\log\frac{z}{1+\sqrt{1+z^2}},\\
t(z)=&\frac{1}{\sqrt{1+z^2}},\\
u_1(t)=&\frac{t}{8}-\frac{5t^3}{24},
\end{split}\end{equation}
one can check that the term $\mathscr{B}_i$ has order $\vep$, and hence would not contribute to the small separation asymptotic expansion up to the next-to-leading order term. Hence,
 \begin{equation}\begin{split}
\mathbb{M}_{m_i, m_{i+1}}^{(\pm)}\sim &-  \frac{1}{\omega  \left(K_{m_i-\frac{1}{2}}^2( \omega)+K_{m_i+\frac{1}{2}}^2(  \omega)\right)}
K_{m_i+m_{i+1}}\left(2\omega(1+\vep)\right)\mathbb{I}.
\end{split}\end{equation}This is independent of $\alpha$ and $+$, $-$. Moreover, ignoring terms with order higher than $\vep$, $\mathbb{M}_{m_i, m_{i+1}}^{(\pm)} =\mathbb{M}_{-m_i, -m_{i+1}}^{(\pm)}$. Hence, after replacing summation by corresponding integrations,
\begin{equation}\begin{split}
E_{\text{Cas}}
\sim &\frac{\hbar cH }{  \pi  R^2} \sum_{s=0}^{\infty}\frac{1}{s+1}\int_{0}^{1}\frac{d\tau}{\tau^3}  \int_{0}^{\infty} dm\, m^2
\int_{ -\infty}^{\infty}dn_1\ldots\int_{ -\infty}^{\infty}dn_s\text{tr}\,\left(\mathbb{M}^{(+)}_{m_0m_1}\mathbb{M}^{(+)}_{m_1m_2}\ldots \mathbb{M}^{(+)}_{m_sm_0}\right).
\end{split}\end{equation}
Now
\begin{equation}\begin{split}
\mathbb{M}_{m_i, m_{i+1}}^{(+)}\sim &-  \frac{1}{\omega \displaystyle \left(1+\frac{K_{m_i+\frac{1}{2}}^2(  \omega)}{K_{m_i-\frac{1}{2}}^2( \omega)}\right)}
\frac{K_{m_i+m_{i+1}}\left(2\omega(1+\vep)\right)}{K_{m_i-\frac{1}{2}}^2( \omega)}\mathbb{I}.
\end{split}\end{equation}
Using the Debye asymptotic expansions, one can find the asymptotic expansions for
$$ \frac{K_{m_i+m_{i+1}}\left(2\omega(1+\vep)\right)}{K_{m_i-\frac{1}{2}}^2( \omega)}$$ and $$\frac{1}{  \displaystyle  1+\frac{K_{m_i+\frac{1}{2}}^2(  \omega)}{K_{m_i-\frac{1}{2}}^2( \omega)} }$$separately, up to terms of order $\vep$.
These give an expansion of the form
\begin{equation}\begin{split}
\mathbb{M}_{m_i, m_{i+1}}^{(\pm)}\sim   &-\frac{1}{2}\sqrt{\frac{\tau}{\pi m}}\exp\left(-\frac{2\vep m}{\tau}-\frac{\tau(n_i-n_{i+1})^2}{4m} \right) \left(1+\mathcal{A}_{i,1}+\mathcal{A}_{i,2}\right)\mathbb{I},
\end{split}\end{equation}
where $\mathcal{A}_{i,1}$ and $\mathcal{A}_{i,2}$ are respectively terms of order $\sqrt{\vep}$ and $\vep$. $\mathcal{A}_{i,1}$  is an odd function in $n_i$ and $n_{i+1}$ and hence integrating it over an even function of $n_i$ gives $0$. As a result, the next-to-leading order term in the  small separation asymptotic expansion of the Casimir interaction energy is of order $\vep$ smaller than the leading order term. We have
\begin{equation}\begin{split}
E_{\text{Cas}}
\sim &\frac{2\hbar cH }{ \pi  R^2} \sum_{s=0}^{\infty}\frac{(-1)^{s+1}}{s+1}\frac{1}{2^{s+1}\pi^{\frac{s+1}{2}}}\int_{0}^{1} d\tau  \tau^{\frac{s-5}{2}} \int_{0}^{\infty} dm\, m^{-\frac{s-3}{2}}
\int_{ -\infty}^{\infty}dn_1\ldots\int_{ -\infty}^{\infty}dn_s\\
&\times \exp\left(-\frac{2\vep (s+1)m}{\tau}-\sum_{i=0}^s\frac{\tau(n_i-n_{i+1})^2}{4m} \right)\left(1+\sum_{i=0}^{s-1}\sum_{j=i+1}^s\mathcal{A}_{i,1}\mathcal{A}_{j,1}+\sum_{i=0}^s\mathcal{A}_{i,2}  \right).\end{split}\end{equation}
The integration over $n_i$ are standard Gaussian integrations and the integration over $m$ and $\tau$ can also be performed explicitly. We find that
\begin{equation}\label{eq3_27_10}\begin{split}
E_{\text{Cas}}  \sim &\frac{3\hbar cH }{ 16\sqrt{2}\pi\vep^{\frac{5}{2}}  R^2} \sum_{s=0}^{\infty}\frac{(-1)^{s+1}}{(s+1)^{4}} \left(1+\left(\frac{7}{36}-\frac{1}{9}(s+1)^2\right)\
  \frac{d}{R} \right)\\
  \sim &-\frac{7\pi^3\hbar cH \sqrt{R}}{ 3840\sqrt{2} d^{\frac{5}{2}}   } \left(1+\left[\frac{7}{36}-\frac{20}{21\pi^2}\right]\frac{d}{R}\right).
\end{split}\end{equation}One can easily check that the leading term
\begin{align}
E_{\text{Cas}}  \sim & -\frac{ 7\pi^3\hbar cH \sqrt{R}}{ 3840\sqrt{2} d^{\frac{5}{2}}   }
\end{align}  coincides with the proximity force approximation.

The corresponding results for Dirichlet (D), Neumann (N) and perfectly conducting (C) boundary conditions have been obtained in \cite{14}:
\begin{equation}\label{eq3_27_7}
\begin{split}
E_{\text{Cas}}^{\text{D}}  \sim & -\frac{ \pi^3\hbar cH \sqrt{R}}{ 1920\sqrt{2} d^{\frac{5}{2}}   } \left(1+ \frac{7}{36} \frac{d}{R}\right),\\
E_{\text{Cas}}^{\text{N}}  \sim & -\frac{ \pi^3\hbar cH \sqrt{R}}{ 1920\sqrt{2} d^{\frac{5}{2}}   } \left(1+\left[\frac{7}{36}-\frac{40}{3\pi^2}\right]\frac{d}{R}\right),\\
E_{\text{Cas}}^{\text{C}}  \sim & -\frac{ \pi^3\hbar cH \sqrt{R}}{ 960\sqrt{2} d^{\frac{5}{2}}   } \left(1+\left[\frac{7}{36}-\frac{20}{3\pi^2}\right]\frac{d}{R}\right).
\end{split}
\end{equation}

\subsection{The case of two cylinders outside each other}
In the case that two parallel cylinders of radii $R_A$ and $R_B$ are outside each other,
let
\begin{gather*}
\omega=(R_A+R_B)u,\quad \vep=\frac{d}{R_A+R_B}=\frac{L}{R}-1,\\
a=\frac{R_A}{R_A+R_B},\quad b=\frac{R_B}{R_A+R_B}.
\end{gather*}
Then
\begin{equation}\begin{split}
\mathbb{M}_{m_i,m_{i+1}}^{(\pm)}=& -\frac{1}{  a\omega \sin\alpha \left(K_{m_i-\frac{1}{2}}^2(a\omega)+K_{m_i+\frac{1}{2}}^2(a\omega)\right)}\sum_{m_i'}\frac{K_{m_i'+m_i}\left( (1+\vep)\omega\right)K_{m_i'+m_{i+1}}\left((1+\vep)\omega\right)}{  b\omega\sin\alpha \left(K_{m_i'-\frac{1}{2}}^2(b\omega)+K_{m_i'+\frac{1}{2}}^2(b\omega)\right)}\\
&\times \begin{pmatrix}-1\mp i \mathscr{B}_i\sin\alpha & i\cos\alpha \\ i\cos\alpha& 1\mp i\mathscr{B}_i\sin\alpha \end{pmatrix} \begin{pmatrix}-1\pm i \mathscr{C}_i\sin\alpha & i\cos\alpha \\ i\cos\alpha& 1\pm i\mathscr{C}_i\sin\alpha \end{pmatrix},
\end{split}\end{equation}where
\begin{equation}\begin{split}
 \mathscr{B}_i=&a\omega  \left(K_{m_i-\frac{1}{2}}(a\omega)I_{m_i-\frac{1}{2} }(a\omega)- K_{m_i+\frac{1}{2}}(a\omega)I_{m_i+\frac{1}{2}}(a\omega)\right),\\
\mathscr{C}_i=&b\omega \left(K_{m_i'-\frac{1}{2}}(b\omega)I_{m_i'-\frac{1}{2}}(b\omega)- K_{m_i'+\frac{1}{2}}(b\omega)I_{m_i'+\frac{1}{2}}(b\omega)\right).
\end{split}\end{equation}Let
\begin{gather*}m_0=m,\quad m_i=m+n_i,\\ m_i'=\frac{b}{2a}(m_i+m_{i+1})+q_i=\frac{b}{a}m+\frac{b}{2a}(n_i+n_{i+1})+q_i,\\
\omega=\frac{m\sqrt{1-\tau^2}}{a\tau}.
\end{gather*}As before, $m$ has order $\vep^{-1}$, $n_i$ and $q_i$ has order $\vep^{-\frac{1}{2}}$.

Now,
\begin{equation}\begin{split}
\begin{pmatrix}-1\mp i \mathscr{B}_i\sin\alpha & i\cos\alpha \\ i\cos\alpha& 1\mp i\mathscr{B}_i\sin\alpha \end{pmatrix} \begin{pmatrix}-1\pm i \mathscr{C}_i\sin\alpha & i\cos\alpha \\ i\cos\alpha& 1\pm i\mathscr{C}_i\sin\alpha \end{pmatrix}= &\begin{pmatrix}\sin^2\alpha\pm i \left(\mathscr{B}_i-\mathscr{C}_i\right)\sin\alpha &\pm  \left(\mathscr{B}_i-\mathscr{C}_i\right)\sin\alpha\cos\alpha \\ \pm  \left(\mathscr{B}_i-\mathscr{C}_i\right)\sin\alpha\cos\alpha& \sin^2\alpha\mp i\left(\mathscr{B}_i-\mathscr{C}_i\right)\sin\alpha \end{pmatrix}.
\end{split}\end{equation}
In the same way, we find that the $\mathscr{B}_i$ and $\mathscr{C}_i$ terms would not contribute to the leading and next-to-leading order terms of the Casimir interaction energy.

Hence,
\begin{equation}\begin{split}
E_{\text{Cas}}
\sim & \frac{\hbar cH }{  \pi R_A^2}\sum_{s=0}^{\infty} \frac{1}{s+1}\int_{0}^{1} \frac{d\tau}{\tau^3}\int_0^{\infty} dm m^2  \int_{-\infty}^{\infty} dn_1\ldots\int_{-\infty}^{\infty} dn_s
\text{tr}\,\left( \mathbb{M}^{(+)}_{m_0,m_1}\ldots\mathbb{M}^{(+)}_{m_s,m_0}\right),
\end{split}\end{equation}where
\begin{equation}\begin{split}
\mathbb{M}_{m_i,m_{i+1}}\sim & -\frac{1}{ a\omega  \left(K_{m_i-\frac{1}{2}}^2(a\omega)+K_{m_i+\frac{1}{2}}^2(a\omega)\right)}\int_{-\infty}^{\infty} dq_i \frac{K_{m_i'+m_i}(  (1+\vep)\omega )K_{m_i'+m_{i+1}}((1+\vep)\omega)}{ b\omega  \left(K_{m_i'-\frac{1}{2}}^2(b\omega)+K_{m_i'+\frac{1}{2}}^2(b\omega)\right)}\mathbb{I}.
\end{split}\end{equation}
Using the Debye asymptotic expansions \eqref{eq3_26_6} again, we can find the small $\vep$ expansions for
$$
 \frac{K_{m_i'+m_i}(  (1+\vep)\omega )K_{m_{i+1}+m_i'}((1+\vep)\omega)}{K_{m_i-\frac{1}{2}}^2(a\omega)K_{m_i'-\frac{1}{2}}^2(b\omega)},
$$$$
\frac{1}{\displaystyle 1+\frac{K_{m_i'+\frac{1}{2}}^2(b\omega)}{K_{m_i'-\frac{1}{2}}^2(b\omega)}}
$$and
$$
\frac{1}{\displaystyle 1+\frac{K_{m_i+\frac{1}{2}}^2(a\omega)}{K_{m_i-\frac{1}{2}}^2(a\omega)}}
$$up to terms of order $\vep$.
These give
\begin{equation}\begin{split}
\mathbb{M}_{m_i,m_{i+1}}\sim & -\frac{a\tau}{2  \pi m}\int_{-\infty}^{\infty}dq_i\exp\left(-\frac{2\vep m}{a\tau}-\frac{b\tau(n_i-n_{i+1})^2}{4m}-\frac{a^2\tau}{bm}q_i^2 \right)\left(1+\mathcal{B}_{i,1}+\mathcal{B}_{i,2}\right) \mathbb{I}.\end{split}\end{equation}
After the Gaussian integration over $q_i$, we have
\begin{equation}\begin{split}
\mathbb{M}_{m_i,m_{i+1}}\sim & -\frac{\sqrt{b\tau}}{2  \sqrt{\pi m}} \exp\left(-\frac{2\vep m}{a\tau}-\frac{b\tau(n_i-n_{i+1})^2}{4m} \right)\left(1+\mathcal{C}_{i,1}+\mathcal{C}_{i,2}\right)
\mathbb{I}.
\end{split}\end{equation}
Hence,
\begin{equation}\begin{split}
E_{\text{Cas}}
\sim & \frac{2\hbar cH }{  \pi R_A^2}\sum_{s=0}^{\infty} \frac{(-1)^{s+1}}{s+1}\frac{b^{\frac{s+1}{2}}}{2^{s+1}\pi^{\frac{s+1}{2}}}\int_{0}^{1}  d\tau \tau^{\frac{s-5}{2}}\int_0^{\infty} dm m^{-\frac{s-3}{2}}  \int_{-\infty}^{\infty} dn_1\ldots\int_{-\infty}^{\infty} dn_s
\\&\times\exp\left(-\frac{2\vep (s+1) m}{a\tau}-\sum_{i=0}^s\frac{b\tau(n_i-n_{i+1})^2}{4m} \right)
 \left(1+\sum_{i=0}^{s-1}\sum_{j=i+1}^s\mathcal{C}_{i,1}\mathcal{C}_{j,1}+\sum_{i=0}^{s}\mathcal{C}_{i,2}\right).\end{split}\end{equation}
As before, integrations over $n_i$, $m$ and $\tau$ give
 \begin{equation}\label{eq3_27_13}\begin{split}
E_{\text{Cas}}\sim & \frac{3\hbar cH a^{\frac{5}{2}}\sqrt{b}}{ 16\sqrt{2} \pi  \vep^{\frac{5}{2}}R_A^2}\sum_{s=0}^{\infty} \frac{(-1)^{s+1}}{(s+1)^{4}} \left(1-\frac{7}{12}\frac{d}{R_A+R_B}+\left[\frac{7}{36}-\frac{(s+1)^2}{9}\right]\left(\frac{d}{R_A}+\frac{d}{R_B}\right)\right)\\
 =&-\frac{7\pi^3\hbar cH \sqrt{R_AR_B}}{ 3840\sqrt{2} d^{\frac{5}{2}} \sqrt{R_A+R_B}  } \left(1-\frac{7}{12}\frac{d}{R_A+R_B}+\left[\frac{7}{36}-\frac{20}{21\pi^2}\right]\left(\frac{d}{R_A}+\frac{d}{R_B}\right)\right).
\end{split}\end{equation}
The leading term
\begin{equation}
E_{\text{Cas}}\sim  -\frac{7\pi^3\hbar cH \sqrt{R_AR_B}}{ 3840\sqrt{2} d^{\frac{5}{2}} \sqrt{R_A+R_B}  }
\end{equation}also coincides with the proximity force approximation.

The corresponding results for Dirichlet, Neumann and perfectly conducting boundary conditions were obtained in \cite{12}. They are given by
\begin{equation}\label{eq3_27_12}
\begin{split}
E_{\text{Cas}}^{\text{D}}\sim & -\frac{\pi^3\hbar cH \sqrt{R_AR_B}}{ 1920\sqrt{2} d^{\frac{5}{2}} \sqrt{R_A+R_B}  } \left(1-\frac{7}{12}\frac{d}{R_A+R_B}+ \frac{7}{36} \left(\frac{d}{R_A}+\frac{d}{R_B}\right)\right),\\
E_{\text{Cas}}^{\text{N}}\sim & -\frac{\pi^3\hbar cH \sqrt{R_AR_B}}{ 1920\sqrt{2} d^{\frac{5}{2}} \sqrt{R_A+R_B}  } \left(1-\frac{7}{12}\frac{d}{R_A+R_B}+ \left[\frac{7}{36}-\frac{40}{3\pi^2}\right] \left(\frac{d}{R_A}+\frac{d}{R_B}\right)\right),\\
E_{\text{Cas}}^{\text{C}}\sim & -\frac{\pi^3\hbar cH \sqrt{R_AR_B}}{ 960\sqrt{2} d^{\frac{5}{2}} \sqrt{R_A+R_B}  } \left(1-\frac{7}{12}\frac{d}{R_A+R_B}+ \left[\frac{7}{36}-\frac{20}{3\pi^2}\right] \left(\frac{d}{R_A}+\frac{d}{R_B}\right)\right).
\end{split}
\end{equation}
Notice that for all different boundary conditions, the ratio of the next-to-leading order term to the leading order term contains the universal terms
$$-\frac{7}{12}\frac{d}{R_A+R_B}$$ and $$\frac{7}{36} \left(\frac{d}{R_A}+\frac{d}{R_B}\right).$$

\subsection{The case of one cylinder inside another}
 In the case that the cylinder of radius $R_A$ is inside the cylinder of radius $R_B$,
let
\begin{gather*}
\omega=(R_B-R_A)u=\frac{m\sqrt{1-\tau^2}}{a\tau},\\
\vep=\frac{d}{R_B-R_A}=1-\frac{L}{R},\\
a=\frac{R_A}{R_B-R_A},\quad b=\frac{R_B}{R_B-R_A},\\
m_i=m+n_i,\quad m_i'=\frac{b}{2a}(n_i+n_{i+1})+q_i=\frac{b}{a}m+\frac{b}{2a}(\tilde{n}_i+\tilde{n}_{i+1})+q_i.
\end{gather*}
Then as in the case of two cylinders outside each other, we find that
\begin{equation}\begin{split}
\mathbb{M}_{m_i,m_{i+1}}^{(\pm)}\sim & -\frac{1}{  a\omega \left(K_{m_i-\frac{1}{2}}^2(a\omega)+K_{m_i+\frac{1}{2}}^2(a\omega)\right)}\int_{-\infty}^{\infty}dq_i\frac{I_{m_i'-m_i}\left( (1-\vep)\omega\right)I_{m_i'-m_{i+1}}\left((1-\vep)\omega\right)}{  b\omega\left(I_{m_i'-\frac{1}{2}}^2(b\omega)+I_{m_i'+\frac{1}{2}}^2(b\omega)\right)}\mathbb{I}.
\end{split}\end{equation}In the same way, we have
\begin{equation}\begin{split}
E_{\text{Cas}}
\sim & \frac{3\hbar cH a^{\frac{5}{2}}\sqrt{b}}{ 16\sqrt{2} \pi  \vep^{\frac{5}{2}}R_A^2}\sum_{s=0}^{\infty} \frac{(-1)^{s+1}}{(s+1)^{4}} \left(1+\frac{7}{12}\frac{d}{R_B-R_A}+\left[\frac{7}{36}-\frac{(s+1)^2}{9}\right]\left(\frac{d}{R_A}-\frac{d}{R_B}\right)\right)\\
= &-\frac{7\pi^3\hbar cH \sqrt{R_AR_B}}{ 3840\sqrt{2} d^{\frac{5}{2}} \sqrt{R_B-R_A}  } \left(1+\frac{7}{12}\frac{d}{R_B-R_A}+\left[\frac{7}{36}-\frac{20}{21\pi^2}\right]\left(\frac{d}{R_A}-\frac{d}{R_B}\right)\right).
\end{split}\end{equation}As computed in \cite{12},  the corresponding results for Dirichlet, Neumann and perfectly conducting boundary conditions are given by
\begin{equation}
\begin{split}
E_{\text{Cas}}^{\text{D}}\sim & -\frac{\pi^3\hbar cH \sqrt{R_AR_B}}{ 1920\sqrt{2} d^{\frac{5}{2}} \sqrt{R_B-R_A}  } \left(1+\frac{7}{12}\frac{d}{R_B-R_A}+ \frac{7}{36} \left(\frac{d}{R_A}-\frac{d}{R_B}\right)\right),\\
E_{\text{Cas}}^{\text{N}}\sim & -\frac{\pi^3\hbar cH \sqrt{R_AR_B}}{ 1920\sqrt{2} d^{\frac{5}{2}} \sqrt{R_B-R_A}  } \left(1+\frac{7}{12}\frac{d}{R_B-R_A}+ \left[\frac{7}{36}-\frac{40}{3\pi^2}\right] \left(\frac{d}{R_A}-\frac{d}{R_B}\right)\right),\\
E_{\text{Cas}}^{\text{C}}\sim & -\frac{\pi^3\hbar cH \sqrt{R_AR_B}}{ 960\sqrt{2} d^{\frac{5}{2}} \sqrt{R_B-R_A}  } \left(1+\frac{7}{12}\frac{d}{R_B-R_A}+ \left[\frac{7}{36}-\frac{20}{3\pi^2}\right] \left(\frac{d}{R_A}-\frac{d}{R_B}\right)\right).
\end{split}
\end{equation}
Again, we notice that for all different boundary conditions, the ratio of the next-to-leading order term to the leading order term contains the universal terms
$$\frac{7}{12}\frac{d}{R_B-R_A}$$ and $$\frac{7}{36} \left(\frac{d}{R_A}-\frac{d}{R_B}\right).$$

\section{Discussions}
In the previous section, we have seen that the leading order terms of the Casimir interaction energies are  indeed equal to those derived using proximity force approximation. The asymptotic expansion of the Casimir interaction energy up to the next-to-leading order term has been a subject of much interest. In \cite{61, 62, 63}, Fosco et al performed derivative expansion to the path integral representation of the Casimir energy and obtained an integral expression for the expansion of the Casimir energy up to the next-to-leading order which is not completely convergent. However, it is good enough for computing the next-to-leading order term. Their method works successfully for scalar field but only partially for electromagnetic field. No results have been derived so far for fermionic fields.

Inspired by the work \cite{61}, Bimonte et al   proposed in \cite{60} that the Casimir interaction energy has a derivative expansion of the form
 \begin{equation}\label{eq3_27_5}\begin{split}
 E_{\text{Cas}}^{\text{DE}}=&\int_{\Sigma}d^2\mathbf{x}\, \mathcal{E}_{\text{Cas}}^{\parallel}(H)\Bigl(1+\beta_1(H)\nabla H_1\cdot\nabla H_1+\beta_2(H)\nabla H_2\cdot\nabla H_2\\&\hspace{3cm}+\beta_{\times}(H)\nabla H_1\cdot \nabla H_2+\beta_-(H)\hat{\mathbf{z}}\cdot (\nabla H_1\times \nabla H_2)+\ldots\Bigr),
 \end{split}\end{equation}where $\mathcal{E}_{\text{Cas}}^{\parallel}$ is the Casimir energy density between two parallel plates, $\Sigma$ can be taken to be the $z=0$ plane parametrized by $\mathbf{x}=(x,y)$,   $z=H_1(\mathbf{x})$ and $z=H_2(\mathbf{x})$ are the height profiles of the two objects with respect to $\Sigma$, and $H=H_1-H_2$ is the height difference. The leading term
 $$\int_{\Sigma}d^2\mathbf{x}\, \mathcal{E}_{\text{Cas}}^{\parallel}(H)$$ is precisely the proximity force approximation.
 Using the invariance of the Casimir interaction energy with respect to tilting the reference plane $\Sigma$, it was found that
 \begin{equation*}
 \begin{split}
 \beta_-(H)=&0,\\
 \beta_{\times}(H)=&2-\beta_1(H)-\beta_2(H).
 \end{split}
 \end{equation*}
For Dirichlet, Neumann and perfectly conducting boundary conditions,   $\beta=\beta_1=\beta_2$ is found to be a pure number  that only depends on the boundary conditions, which is given by
\begin{equation}\label{eq3_27_9}
\begin{split}
\beta^{\text{D}}=&\frac{2}{3},\\
\beta^{\text{N}}=&\frac{2}{3}\left(1-\frac{30}{\pi^2}\right),\\
\beta^{\text{C}}=&\frac{2}{3}\left(1-\frac{15}{\pi^2}\right).
\end{split}
\end{equation}
Since
\begin{align}
\mathcal{E}_{\text{Cas}}^{\parallel}(H)=\frac{\alpha}{H^3},
\end{align}where $\alpha$ depends only on boundary conditions, if one let
\begin{equation}\begin{split}
c_0=&\int_{\Sigma}d^2\mathbf{x}\frac{1}{H^3}\\
c_{11}=&\int_{\Sigma}d^2\mathbf{x}\frac{\nabla H_1\cdot\nabla H_1}{H^3},\\
c_{22}=&\int_{\Sigma}d^2\mathbf{x}\frac{\nabla H_2\cdot\nabla H_2}{H^3},\\
c_{12}=&\int_{\Sigma}d^2\mathbf{x}\frac{\nabla H_1\cdot\nabla H_2}{H^3},
\end{split}\end{equation}
then the ansatz \eqref{eq3_27_5} says that
\begin{equation}\label{eq3_27_11}
E_{\text{Cas}}^{\text{DE}}\sim \alpha\Bigl\{c_0+\beta (c_{11}+ c_{22}-2c_{12})+2c_{12}\Bigr\}.
\end{equation}
For   cylinder-plate configuration,  one finds that
\begin{equation}
\begin{split}
c_0=&\frac{3\pi\sqrt{ R}H}{4\sqrt{2}d^{\frac{5}{2}}}\left(1-\frac{d}{4R}+\ldots\right)\\
c_{11}=&\frac{3\pi\sqrt{ R}H}{4\sqrt{2}d^{\frac{5}{2}}}\times \frac{2}{3}\frac{d}{R}+\ldots,\\
c_{22}=&c_{12}=0.
\end{split}\end{equation}
Therefore, the ansatz \eqref{eq3_27_5} says that
\begin{align}\label{eq3_27_8}
E_{\text{Cas}}^{\text{DE}}\sim \frac{3\pi\sqrt{ R}H}{4\sqrt{2}d^{\frac{5}{2}}}\alpha\left\{1+\left(\frac{2}{3}\beta-\frac{1}{4}\right)\frac{d}{R}+\ldots\right\}.
\end{align}As observed in \cite{60}, the results from exact computation \eqref{eq3_27_7} do indeed satisfy \eqref{eq3_27_8} with the various $\beta$ given in  \eqref{eq3_27_9}. Our new result
 \eqref{eq3_27_10} satisfies \eqref{eq3_27_8} if the value of $\beta$ for fermionic fields with MIT bag boundary conditions is given by
\begin{align}\label{eq3_27_15}
\beta^{\text{F}}=\frac{2}{3}-\frac{10}{7\pi^2}.
\end{align}
For two cylinders outside each other,
\begin{equation}\begin{split}
c_0=&\frac{3\pi H}{4\sqrt{2}d^{5/2}}  \frac{\sqrt{R_A R_B}}{\sqrt{ R_A+R_B}}\left(1 +\frac{3}{4}\frac{d}{R_A+R_B}-\frac{ 1 }{4   } \left(\frac{d}{R_A}+\frac{d}{R_B}\right)+\ldots\right),\\
c_{11}=&\frac{3\pi H}{4\sqrt{2}d^{5/2}}  \frac{\sqrt{R_A R_B}}{\sqrt{ R_A+R_B}}\left( -\frac{2}{3}\frac{d}{R_A+R_B}+\frac{ 2 }{3   }  \frac{d}{R_A}  +\ldots\right),\\
c_{22}=&\frac{3\pi H}{4\sqrt{2}d^{5/2}}  \frac{\sqrt{R_A R_B}}{\sqrt{ R_A+R_B}}\left( -\frac{2}{3}\frac{d}{R_A+R_B}+\frac{ 2 }{3   }  \frac{d}{R_B}  +\ldots\right),\\
c_{12}=&\frac{3\pi H}{4\sqrt{2}d^{5/2}}  \frac{\sqrt{R_A R_B}}{\sqrt{ R_A+R_B}}\left(-\frac{2}{3}\frac{d}{R_A+R_B} +\ldots\right).
\end{split}\end{equation}
Hence,
\eqref{eq3_27_11} gives
\begin{align}\label{eq3_27_14}
E_{\text{Cas}}^{\text{DE}}\sim \frac{3\pi\sqrt{ R_AR_B}H}{4\sqrt{2}d^{\frac{5}{2}}\sqrt{ R_A+R_B}}\alpha\left\{1-\frac{7}{12}\frac{d}{R_A+R_B}+\left[\frac{2}{3}\beta-\frac{1}{4}\right]\left(\frac{d}{R_A}+\frac{d}{R_B}\right)+\ldots\right\}.
\end{align}
 One can check immediately that our new result \eqref{eq3_27_13} satisfies \eqref{eq3_27_14} with $\beta_F$ given by \eqref{eq3_27_15}.

A similar computation for two spheres also shows that our results in \cite{59} satisfies \eqref{eq3_27_11} with $\beta_F$ given by \eqref{eq3_27_15}.
\bigskip
\begin{acknowledgments}\noindent
  This work is supported by the Ministry of   Education of Malaysia  under   FRGS grant FRGS/1/2013/ST02/UNIM/02/2.
\end{acknowledgments}

\end{document}